\documentclass[ICDM]{IEEEtran}

\usepackage{multirow}
\ifCLASSINFOpdf
 \usepackage[pdftex]{graphicx}
\else
\fi

\usepackage[normalem]{ulem}
\useunder{\uline}{\ul}{}
\usepackage {booktabs}

\begin{document}

\title{Video Surveillance System Incorporating Expert Decision-making Process: A Case Study on Detecting Calving Signs in Cattle}

\author{%
\IEEEauthorblockN{%
Ryosuke Hyodo\IEEEauthorrefmark{1},
Susumu Saito\IEEEauthorrefmark{1}\IEEEauthorrefmark{2},
Teppei Nakano\IEEEauthorrefmark{1}\IEEEauthorrefmark{2},
Makoto Akabane\IEEEauthorrefmark{1},
Ryoichi Kasuga\IEEEauthorrefmark{3},
Tetsuji Ogawa\IEEEauthorrefmark{1}
}

\IEEEauthorblockA{%
\IEEEauthorrefmark{1}
Waseda University, Tokyo, Japan \\
}
\IEEEauthorblockA{%
\IEEEauthorrefmark{2}
Intelligent Framework Lab, Tokyo, Japan\\
}
\IEEEauthorblockA{%
\IEEEauthorrefmark{3}
Farmers Support, Kagoshima, Japan\\
}

%
% \authorblockA{
% Email: \{yaegashi, susumu, teppei, ogawa\}@pcl.cs.waseda.ac.jp\\}
}

\maketitle

\begin{abstract}
Through a user study in the field of livestock farming,
we verify the effectiveness of 
an XAI framework 
for video surveillance systems.
The systems can be made interpretable 
by incorporating experts' decision-making processes.
AI systems are becoming increasingly common in real-world applications, 
especially in fields related to human decision-making, 
and its interpretability is necessary.
However, there are still relatively few standard methods 
for assessing and addressing the interpretability 
of machine learning-based systems in real-world applications.
In this study, 
we examine the framework of a video surveillance AI system 
that presents the reasoning behind predictions 
by incorporating experts' decision-making processes
with rich domain knowledge of the notification target.
While general \textit{black-box} AI systems can only present final probability values, 
the proposed framework can present information relevant to experts' decisions, 
which is expected to be more helpful for their decision-making.
In our case study, 
we designed a system for detecting signs of calving in cattle
based on the proposed framework 
and evaluated the system through a user study (N=6) 
with people involved in livestock farming.
A comparison with the \textit{black-box} AI system
revealed that many participants referred to the presented reasons
for the prediction results, 
and five out of six participants selected the proposed system 
as the system they would like to use in the future.
It became clear that 
we need to design a user interface that considers 
the reasons for the prediction results.

\textsl{Index Terms}-XAI, machine learning, user study, precision livestock farming
\end{abstract}

\IEEEpeerreviewmaketitle

\section{Introduction}
In recent years, 
artificial intelligence (AI) and machine learning (ML) 
have been used in areas related to human decision-making,
including the medical field~\cite{Emma2020}.
When developing AI systems,
an \textit{end-to-end} approach is generally considered to be 
an indispensable technique due to its simplicity.
However, it has been pointed out that 
this approach is \textit{black-box} in its internal behavior 
and is not a sufficient system for supporting human decision-making~\cite{Abdul2018}.
Meanwhile, 
there has been growing interest in the interpretability of AI systems 
as explainable AI (XAI)~\cite{doshivelez2017,lipton2017}.
Representative technologies of XAI include 
visualization techniques such as CAM~\cite{Zhou2016} and Grad-CAM~\cite{Selvaraju2019}, 
and post-hoc explanations such as local interpretable model agnostic explanations (LIME)~\cite{Marco2016}
and Shapley additive explanations (SHAP)~\cite{Scott2017}.
However, many developments in AI and ML tend to 
suffer from a lack of usability and 
practical interpretability for real decision-makers~\cite{doshivelez2017,lipton2017,miller2018}.

In recent years, 
the human-computer interaction (HCI) community has recognized the importance of human-centered evaluation, 
which incorporates user evaluation into the interpretability of AI systems.
There has been increasing research 
on the user evaluation of these explanatory techniques using experimental data sets~\cite{Harmanpreet2020,Abdul2020,Yin2019}.
However, when implementing AI systems 
for highly specialized tasks in industry, 
the aforementioned generic explanatory techniques 
and the findings on general user evaluations using experimental datasets 
do not lead to interpretability for experts.
Thus, 
system design and empirical experiments in highly specialized tasks are essential;
however, there is still a lack of research on XAI frameworks for industrial use 
and their validation through user studies~\cite{Wang2019}.

This study examines the framework of a video surveillance system 
that can provide relevant explanations for expert judgments 
by incorporating their decision-making processes into the neural networks.
The experts watch a video and 
determine whether there are any abnormalities 
after considering various attributes based on their domain knowledge. 
We design a neural network to incorporate this process into anomaly detection.
Specifically, 
we extract different features
related to an anomaly based on domain knowledge, 
and each stream with extracted features as input
determine whether the features are anomalous.
Here, 
explicitly extracting the features that experts use to make judgments 
enables them to be presented as interpretable information for the experts

\begin{figure*}
    \begin{center}
    \includegraphics[width=0.8\linewidth]{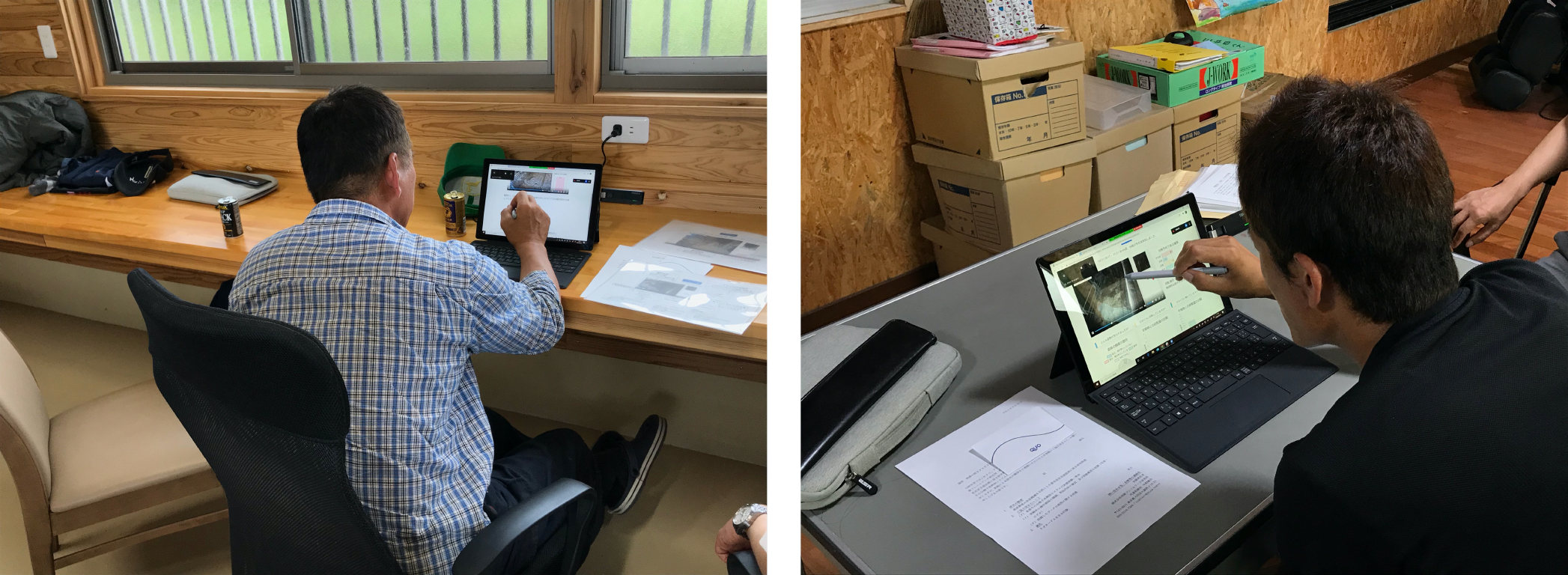}
    \caption{User study of livestock farmers.}
    \label{fig:exp_photo}
    \end{center}
\end{figure*}

The above framework is applied 
to video-based detection for calving signs in cattle 
thorough a user study with livestock farmers (Fig.~\ref{fig:exp_photo}).
Since calf deaths during calving can be very costly for farmers~\cite{McGuirk685}, 
accurately detecting the signs of calving is important for livestock management.
Although contact-type sensors,
which are attached directly to cattle,
are widely used, 
a detection system using a camera,
which is a non-contact-type sensor, 
is valuable in terms of management efficiency and animal welfare.
First, 
to apply the above framework, 
we reviewed the literature on animal studies and conducted farmer interviews 
to collect domain knowledge about calving signs.
Using our findings, 
we designed a system that explicitly extracts statistical information 
on the posture, rotation, and movement of cattle relevant to calving
and then identifies the signs from these features.
We designed the user interface of the notification screen 
using the information provided by the feature extractor.
In the user interface, 
the frequency of posture, 
amount of rotation, 
amount of movement, 
and the pattern of movement (bird's eye view) of cattle are presented to farmers.
Compared with the user interface of the end-to-end system,
which only presents the probability values of calving signs, 
the presentation of the internal state of the system
is expected to be more interpretable for farmers.
The user interface of the end-to-end system and the proposed system 
were evaluated by people who were involved in livestock farming (N=6).
We anticipate that our findings will be useful in developing 
human-centered explainable AI-based systems 
that effectively incorporate experts’ knowledge.

The rest of the paper is organized as follows. 
In Section \ref{sec:methodology},
we briefly explain the framework 
which enables us to understand 
the reasoning behind predictions.
Section \ref{sec:casestudy} describes 
the case study on a calving detection system in livestock farming based on our framework.
In Section \ref{sec:discussion}, 
we discuss the findings from the case study and future work of the proposed framework.
We conclude in Section \ref{sec:conclusion} with a brief summary.

\section{Framework}
\label{sec:methodology}
The proposed framework incorporates 
the users' decision-making process 
into a prediction network 
and allows for interpretation on the basis of their domain knowledge.
A typical machine leaning-based system
uses an end-to-end approach with experts' annotations 
to model the notification target.
In contrast,
the proposed framework uses a human-centric approach 
and models the notification target through user interviews.
This section describes 
the following four phases of the design procedure for the proposed framework:
%%------------------------------
\begin{itemize}
    \item Interview with an expert
    \item Designing streams to extract the information that characterizes what is being monitored
    \item Designing detectors by integrating information from multiple streams
    \item Designing the notification interface
\end{itemize}
%%------------------------------

% 1. 専門家へのインタビュー
In the first phase, 
we interview experts on the notification target.
In interviews,
we ask them to verbalize their decision-making processes 
when detecting the target, 
and determine the features they focus on in daily operations.
Then we divide those features into 
more detailed features that can be judged even by non-experts.

% 2. 抽出された特徴毎に、その特徴量を抽出するネットワーク（識別機、ストリーム）を構築
In the next phase, 
we extract more detailed features from the ones 
that experts focus on 
when they detect the notification target,
and then we develop a stream network based on the extracted features.
Dividing the features into more detailed features
that can be judged even by non-experts,
enables the addition of crowd-sourced annotations.
This is important in terms of making the surveillance system sustainable, 
as it allows us to increase the amount of training data
without having to rely on experts in the domain.

% 3. 複数の特徴量抽出機のストリームを統合するイベント検知ネットワーク（検知器）を構築
In the third phase, 
we develop a network that integrates multiple stream networks
on the basis of different attribute features.
The users are notified 
based on the posterior probability of the network.
Here, 
even if the detection of each stream network is not 
necessarily effective, 
if the predictions of each stream are complementary, 
the detection can be improved.

% 4. 通知UIを構築
Finally, 
we design the user interface that is presented to the users.
In addition to the final posterior probability of the notification target, 
multiple features extracted from each stream network
can be presented as the reasons for the prediction results.
These features are designed with consideration for the responses in the user interviews, 
so they provide information necessary for users' decision-making processes.

\section{Case Study}
\label{sec:casestudy}
In this section, 
we verify the effectiveness of the proposed framework 
for detecting signs of calving in cattle
through a user study with livestock farmers.
First, we describe the system design based on our framework introduced in Section~\ref{sec:methodology}, 
and then we describe the procedure 
for the user experiments and the results and findings.

\subsection{System Design}
In this section, 
we describe the design of our calving detection system
based on the proposed framework.
We describe the domain knowledge of calving signs 
and the network structure that explicitly extracts features relevant to calving.
Finally, 
we explain the user interface that 
presents the farmers with the reasons for the prediction results.

\subsubsection{Calving Signs Observable from Video}

This section presents an overview of the domain knowledge about calving signs.
We interviewed farmers and reviewed the animal science literature to identify cattle behaviors related to calving.
Changes in postures and behaviors related to calving 
have been extensively investigated in animal science.
This part corresponds to the implementation of the first phase of the proposed framework.

%%%%%%%%%%%%%%%%%%%%%%%%%%%%%%%%%%%%%%%%%%%%%%%%%%%%%%%%%%%%
The following are typical posture-based calving signs that can be observed from images:
%%------------------------------
\begin{itemize}
\item\textbf{Switching between standing and lying postures}: About two to six hours before calving, the number of posture changes (\textit{e.g.,} switching between standing and lying) become more frequent~\cite{Speroni2018, saint-dizier2015, Jensen2012} and the time spent lying increases two hours before calving~\cite{Jensen2012}. 

\item\textbf{Tail raising}: About four to six hours before calving, tail raising becomes more frequent~\cite{Jensen2012} and the position of the tail before calving is elevated~\cite{BUENO1981599, OWENS1985321}. 
\end{itemize} 
%%------------------------------
% In this study, 
% we examine the frequency of a cattle standing, lying,
% and raising its tail.

The following action-based calving signs were observed in the video:
%%------------------------------
\begin{itemize}
\item\textbf{Increase in the number of rotations and turns}: Characteristic walking patterns (\textit{e.g.,} rotations and turns) can be observed four hours before calving and become more frequent three hours before calving~\cite{Sugawara}. 

\item\textbf{Increase in aimless walking time}: The duration of walking on the calving day increases~\cite{saint-dizier2015, Jensen2012}. Aimless walking time apparently increases about 140 minutes before calving~\cite{OWENS1985321}. 
\end{itemize}
%%------------------------------
% The developed system integrates the statistics on the cattle's rotation and movements.

\subsubsection{System Architecture}
We designed a multi-stream network~\cite{Hyodo2020} that 
extracts statistical information on posture, rotation, and movement 
based on calving signs identified in prior studies
and integrates the three streams depending on the situation.
This part corresponds to the implementation 
of the second and third phases of the proposed framework.
% In this study, for simplicity, 
% we define the label as the time from just before birth to 3 hours before calving (a positive case) 
% and the time from 24 to 27 hours before calving (a negative case), 
% and predict the label from the 30 minutes of input video frames. 
Specifically, 
each stream identifies calving signs for each 30-minute input video based on features as follows:
%%------------------------------
\begin{itemize}
    \item {\bf Posture-based feature}:
    The appearance of a cattle standing, lying,
    and raising its tail are captured for each video frame using ResNet-50~\cite{7780459}
    and then accumulated into the relevant frequencies
    using temporal pooling techniques.
    
    \item {\bf Rotation-based feature}:
    Information on body direction is extracted from each video frame using ResNet-50
    and accumulated into a statistic
    on the cattle's rotation by measuring the changes in the body direction using the M-measure~\cite{Hermansky,Ogawa}.
        
    \item {\bf Movement-based feature}:
    The region of the cattle's body is detected in each video frame using YOLOv3~\cite{redmon}
    and differences in locations across frames are accumulated into a statistic on the cattle's movement.
\end{itemize}
%%------------------------------
The calving-relevant features
are designed to be extracted from 
information that can be judged by non-experts, 
such as posture, neck and tail positions,
and positional coordinates.
This makes it possible to collect data using crowdsourcing,
and re-training the feature extraction mechanism becomes easier.

% These features were designed 
% based on the animal science and interviews with farmers on calving signs.

\subsubsection{User Interface of Notification Screen}
In this section, 
we describe the design of the user interface of the proposed system.
This part corresponds to the implementation of the fourth phase of the proposed framework.
In addition to the final posterior probability of calving signs, 
the proposed framework provides 
the following information related to calving signs for each frame:
\begin{itemize}
    \item Posterior probability of cattle's posture, 1) standing cattle with tail raised, 2) standing cattle without tail raised, 3) lying cattle, and 4) can't tell
    \item Heatmaps of cattle's body direction
    \item Position coordinates of cattle
\end{itemize}
Displaying these data in an easy-to-understand representation 
will help farmers to estimate when cattle start calving.

\begin{figure*}
    \centering
    \includegraphics[width=0.7\linewidth]{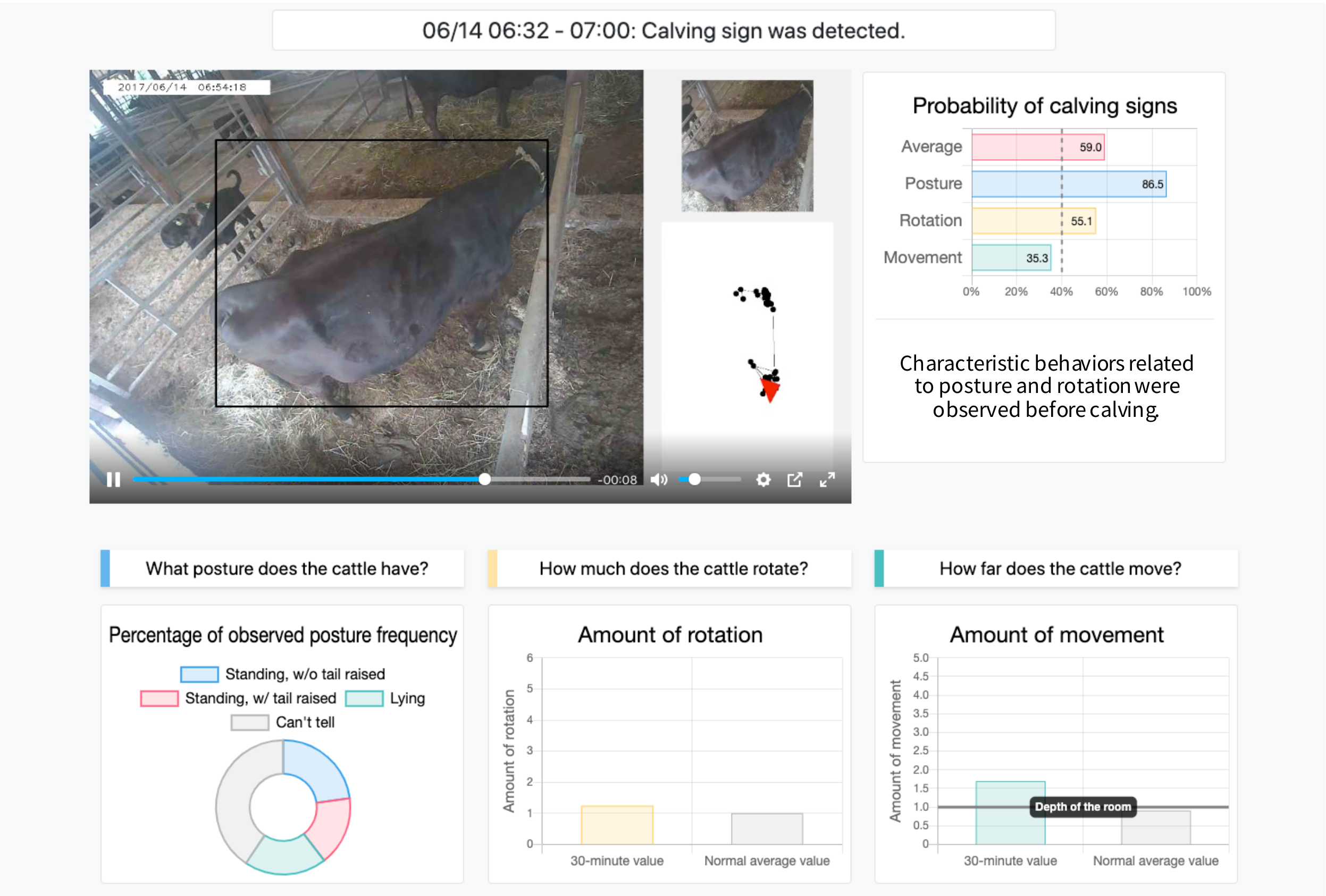}
    \caption{User interface of proposed system (System B). 
             User interface presented to participants was in Japanese.}
    \label{fig:UI_B}
\end{figure*}

Fig.~\ref{fig:UI_B} shows the user interface
designed using the information described above.
In addition to the monitoring video 
and the posterior probability values of the system in this scene, 
the interface also displays 
information on the frequency of posture, 
amount of rotation, 
amount of movement, 
and trajectory of the cattle as seen from directly above.
The upper right bar graph shows 
the posterior probability of calving signs 
which the system output using the 30-minute video frames. 
Four graphs are presented,
each with posterior probabilities and their mean values
based on the statistics on posture, rotation, and movement. 
Users can check which information the system considers to be a calving sign.
The statistics on posture, rotation, and movement are presented
at the bottom of the screen.
The circle graph on the left shows
the frequency of these 30-minute postures
calculated from the posterior probability 
of the posture classification obtained by the feature extractor. 
This graph can be interpreted as 
the reasons for the prediction results of the posture-based stream.
The comparison with the normal state
in the amount of rotation (shown in the center) 
is calculated from the estimated heatmaps 
of the cattle's body direction obtained from the feature extractor.
It is possible to quantify 
how much the cattle turn during these 30 minutes 
compared with their default state, 
which can then be interpreted as reasons for the prediction results. 
In the prototype user interface, 
we visualized the time-series changes in the angle of body direction.
After interviewing a farmer, 
a bar graph was used to simplify the representation.
The comparison with the normal state 
in the amount of movement (shown on the right side)
is calculated from the amount of change in the positional coordinates of the cattle.
Compared with the normal state,
the amount of movement of cattle during this 30-minute
can be quantified and interpreted as the reasons for the prediction results of the movement-based stream.
Finally, 
a bird's eye view of the cattle's position
is displayed on the right side of the monitoring video,
which shows the trajectory of the cattle's position 
as seen from directly above the room.
This allows us to understand 
the general movement pattern 
without continuously watching the video.
%%------------------------------
%%------------------------------

\subsection{User Study}
\begin{figure*}
    \centering
    \includegraphics[width=0.7\linewidth]{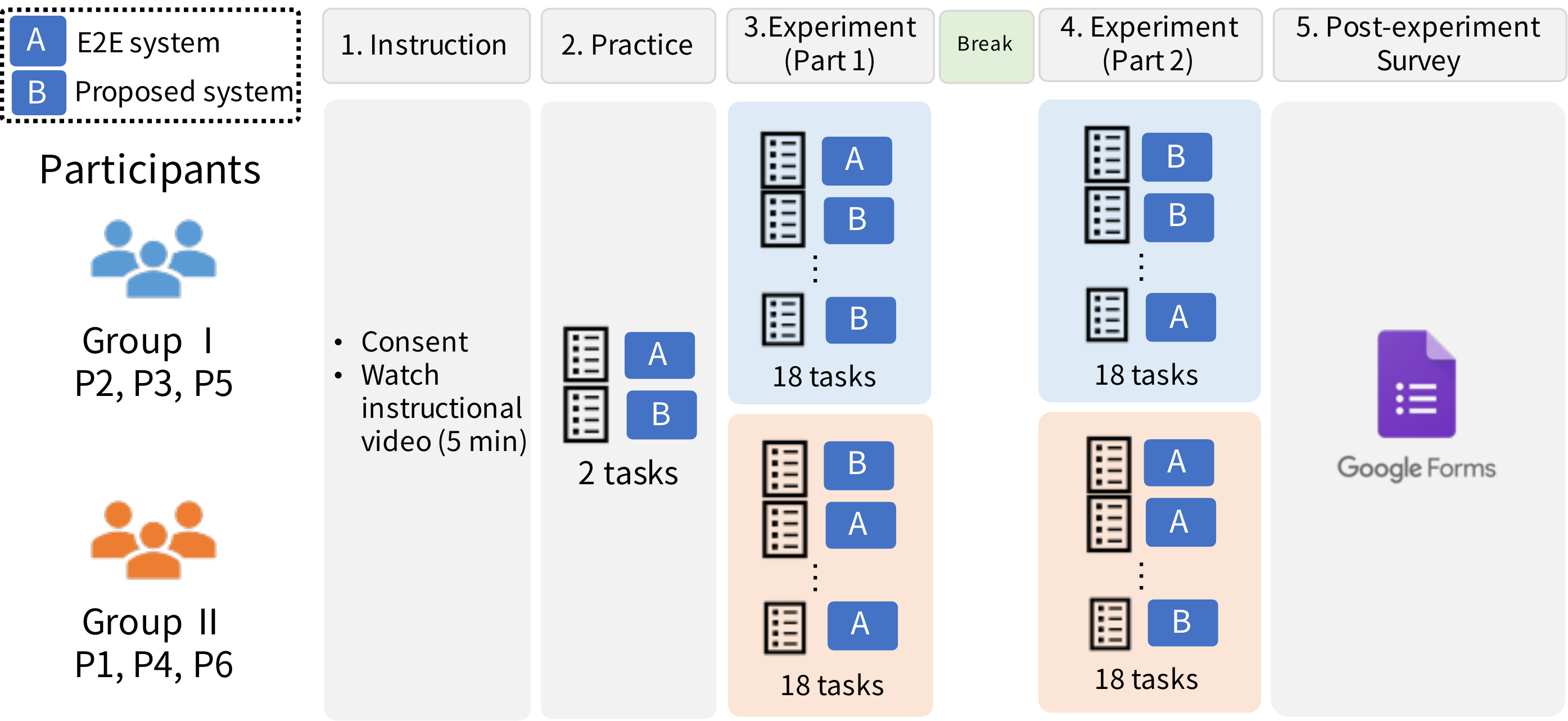}
    \caption{Schematic diagram of experimental procedure. 
            Order of tasks differed for each experimental group.}
    \label{fig:exp_flow}
\end{figure*}

% 実験の内容
To evaluate the proposed framework, 
we experimentally compared it with the end-to-end system were conducted.
% 実験手順
The experiment took about 60-90 minutes in total 
and consisted of instruction, practice, 
experiment part 1, part 2, and a post-experiment survey~(Fig.~\ref{fig:exp_flow}).
First, 
the participants watched a 5-minute instructional video on the experiment.
Afterwards, 
we obtained written informed consent 
signed by the participants before the experiments, 
and the participants were granted a gift card
of ¥ 5,000 JPY (roughly \$50 USD).
As an exercise, 
the participants responded to two notifications 
from the proposed system and the end-to-end system.
In this section, 
the participants were instructed to think aloud~\cite{Monique2004}
and practiced answering questions by actively commenting on
what they saw and thought during the experiment.
The experiment consisted of two parts with a break in between, 
and the participants responded to 36 notifications in total. 
The participants were divided into two experimental groups, 
and each group was shown the user interface 
in a pseudo-random order 
to account for the sequential effects of the interface order. 
After the experiment, 
a qualitative post-experiment survey was conducted using Google Forms.

\begin{figure*}
    \centering
    \includegraphics[width=0.7\linewidth]{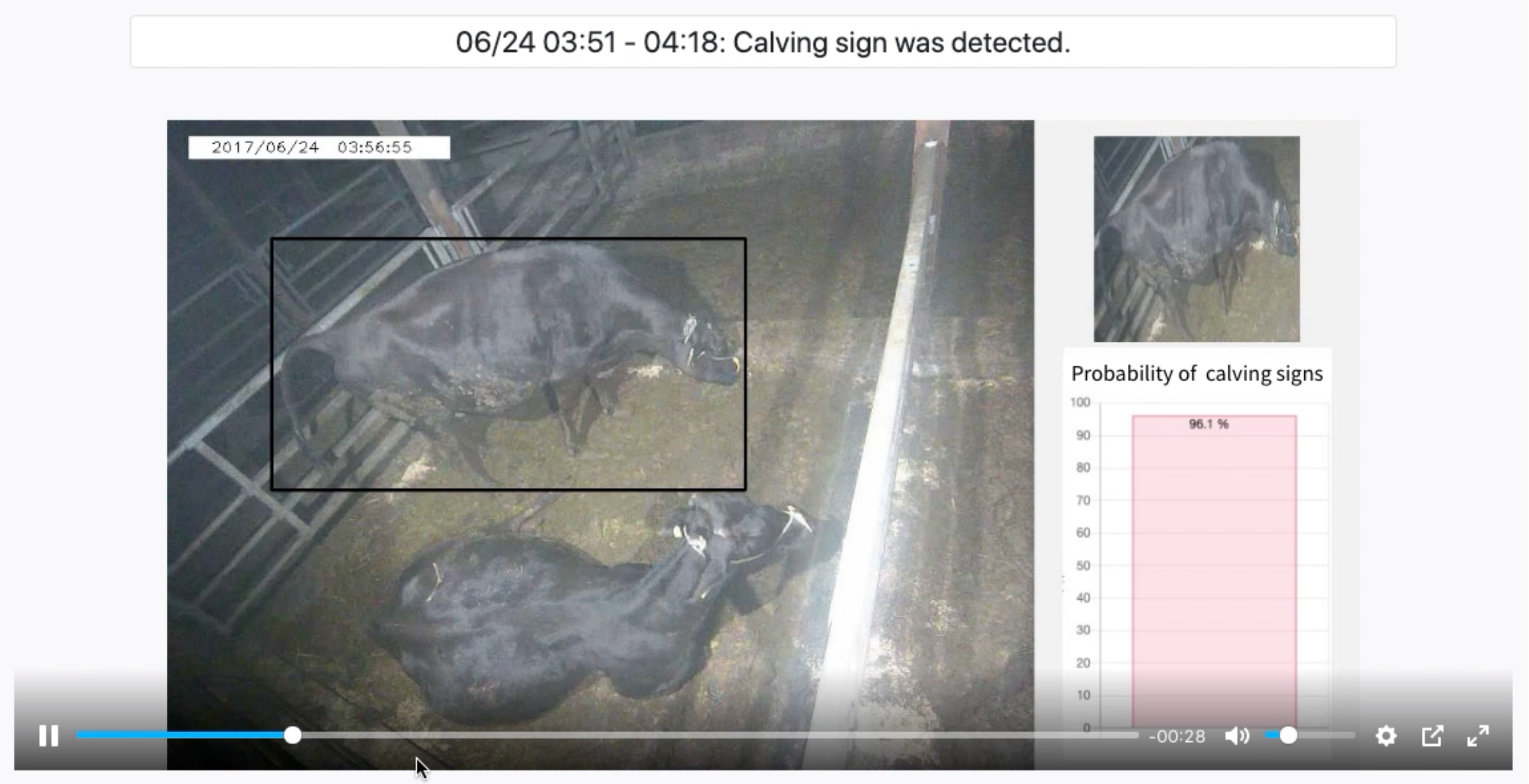}
    \caption{User interface of end-to-end system (System A).
             User interface presented to participants was in Japanese.}
    \label{fig:UI_A}
\end{figure*}

In the experiment, 
the two user interfaces were presented to the participants 
in a pseudo-random order 
on the basis of the experimental group. 
The user interface of 
the end-to-end system for comparison is shown in Fig.~\ref{fig:UI_A}.
Compared with the user interface of the proposed system, 
the end-to-end system presents 
only the posterior probability graph of the prediction result.
For the same 30-minute sequence,
the value of the posterior probability presented 
in the user interface of the end-to-end system 
was the same value as the average posterior probability of each stream
in the user interface of the proposed system.
To avoid bias from the system names, 
we designated the end-to-end system as system A 
and the proposed system as system B
when presenting the user interfaces to the participants.

% 通知単位
The unit of the notification was 
a 30-minute video sequence, 
and a total of 18 sequences were prepared.
A positive case was 
defined as the time from three to zero hours before calving
and a negative case as 
the time from 24 to 27 hours before calving.
The prepared sequences consisted of 
six sequences each of true positives and false positives,
and three sequences each of false negatives and true negatives.
In other words,
we assume that the notification threshold is significantly lowered 
and contains a certain number of false positives.

% 質問事項
The participants are asked to answer 
the following two questions (maximum of four) 
at the bottom of the interface:
\begin{description}
    \item[Q1-1] Did you recognize calving signs in this 30-minutes scene? (-3: Absolutely Not, ... 3: Absolutely Yes)
    \item[Q1-2] What are the reasons for your answer? (Verbal answer)
\end{description}
The participants responded to the following prompts in \textbf{Q1-2}:
1) What did you see on the screen?, 
2) What did you think?, 
and 3) How did you reach your decision?
The intention of this format was 
to reveal as much of their thought process as possible.
After answering the above questions, 
if the answer to \textbf{Q1-1} was ``neither'' or ``predictive'' (0-3), 
the participant was asked to answer the following questions as well:
\begin{description}
    \item[Q2-1] What action would you take after seeing this user interface? (1: Begin assisting immediately, 2: Start making arrangements, 3: Do nothing)
    \item[Q2-2] What are the reasons for your answer? (Verbal answer)
\end{description}
The candidate answers to \textbf{Q2-1} are 
terms referring to common decision-making behaviors for livestock farmers.
The terms are used in the contact-type sensor,
Gyuonkei\footnote{Gyuonkei, http://www.gyuonkei.jp},
a calving notification sensor widely used in Japan. 
Here, ``Begin assisting immediately'' refers to 
the decision to immediately begin assisting with calving, 
and ``Start making arrangements'' refers to 
the preparations made about 24 hours before calving. 
These questions are designed to encourage participants 
to use the interface for their decision-making. 
Each participant is asked to respond to the above questions 
for a total of 36 notifications (2 UIs x 18 sequences).
%%------------------------------
%%-----------------------------
The questions in the post-experiment survey are as follows: 
\begin{description}
    \item[Q'1] Which AI system would you like to use in the future? (1. System A, 2. System B)
    \item[Q'2] Why did you choose that system? (Orally, if you prefer.)
    \item[Q'3] How useful was the information presented in System A? (-2. Useless, ..., 2. Useful)
    \begin{description}
        \item[a] Graph of the probability of calving signs.
    \end{description}
    \item[Q'4]  How useful was the information presented in System B? (-2. Useless, ..., 2. Useful)
    \begin{description}
        \item[a] Graph of the probability of calving signs.
        \item[b] Graph of the posture frequency.
        \item[c] Graph of the amount of rotation.
        \item[d] Graph of the amount of movement.
        \item[e] Bird's eye view (position of the cattle as seen from directly above).
    \end{description}
    \item[Q'5] Which system provided more accurate predictions?(1. System A, 2. System B, 3. Neither)
    \item[Q'6] What are the advantages of the system you chose in \textbf{Q'1} over the other? (Orally, if you prefer.)
    \item[Q'7] What improvements (if any) would you make to System B? (Orally, if you prefer.)
\end{description}

% ユーザの属性
Six people who involved in livestock farming 
participated in the user experiment.
Age, sex, and years of experience in the livestock industry 
of the participants are shown in Table~\ref{tab:profiles}.
The participants belonged to two groups:
farmers (P1-4) and people from an agricultural college (P5-6).
Here, P5 is a professor, and P6 is a student of animal science.
\begin{table}[tb]
\begin{center}
    \caption{Details of experiment participants. 
            Four are farmers~(P1-P4), and other two studies are in agricultural college~(P5-P6).}
    \label{tab:profiles}
    \begin{tabular}{cccccc}
         \toprule
         ID & Group & Age & Sex & Years of experience & \\
         \midrule
         P1 & II & 30's & M & 10-19 \\
         P2 & I & 50's & M & 20-29 \\
         P3 & I & 20's & F & 5-9 \\
         P4 & II & 40's & M & 20-29 \\
         P5 & I & 40's & M & 20-29 \\
         P6 & II & 10's & F & 0-1 \\
         \bottomrule
    \end{tabular}
    \end{center}
\end{table}

% Experimental Setting
The experiment was administered on a computer, 
and the participants accessed the URL to the experiment page provided in advance. 
We used a web conference application (Zoom) 
to record the participants' responses. 
To record the participant's speech during their think-aloud, 
we recorded online on Zoom and locally as a backup.

% \begin{figure}
%     \centering
%     \includegraphics[width=0.7\linewidth]{img/experiments_photo.pdf}
%     \caption{Participants responded by looking at notification screen presented.}
%     \label{fig:exp_photo}
% \end{figure}

\subsection{Results and Findings}
% 回答時の振る舞い
Participants judged 
whether a behavior was a calving sign
from the monitoring video, 
but the behavior of their responses 
varied depending on the type of the user interface presented.
When the interface of the end-to-end system was presented, 
most of the verbal comments were related to the video.
However, when the interface of the proposed system was presented, 
some of the comments were related to the statistics.
One participant shared their judgment, referring to a graph comparing the amount of movement with normal conditions.
% P1 said \textit{"If I look at the graph, the probability of the posture is increasing. In other words, it's not moving."} and 
P5 stated \textit{``I think the cattle is near to calving because the data shows a lot of movement and an increase in rotation.''}.
Given the trend of the responses, 
we believe that presenting the internal state 
of the proposed system 
was effective for judging whether a behavior was a calving sign.

\begin{figure*}
    \centering
    \includegraphics[width=0.9\linewidth]{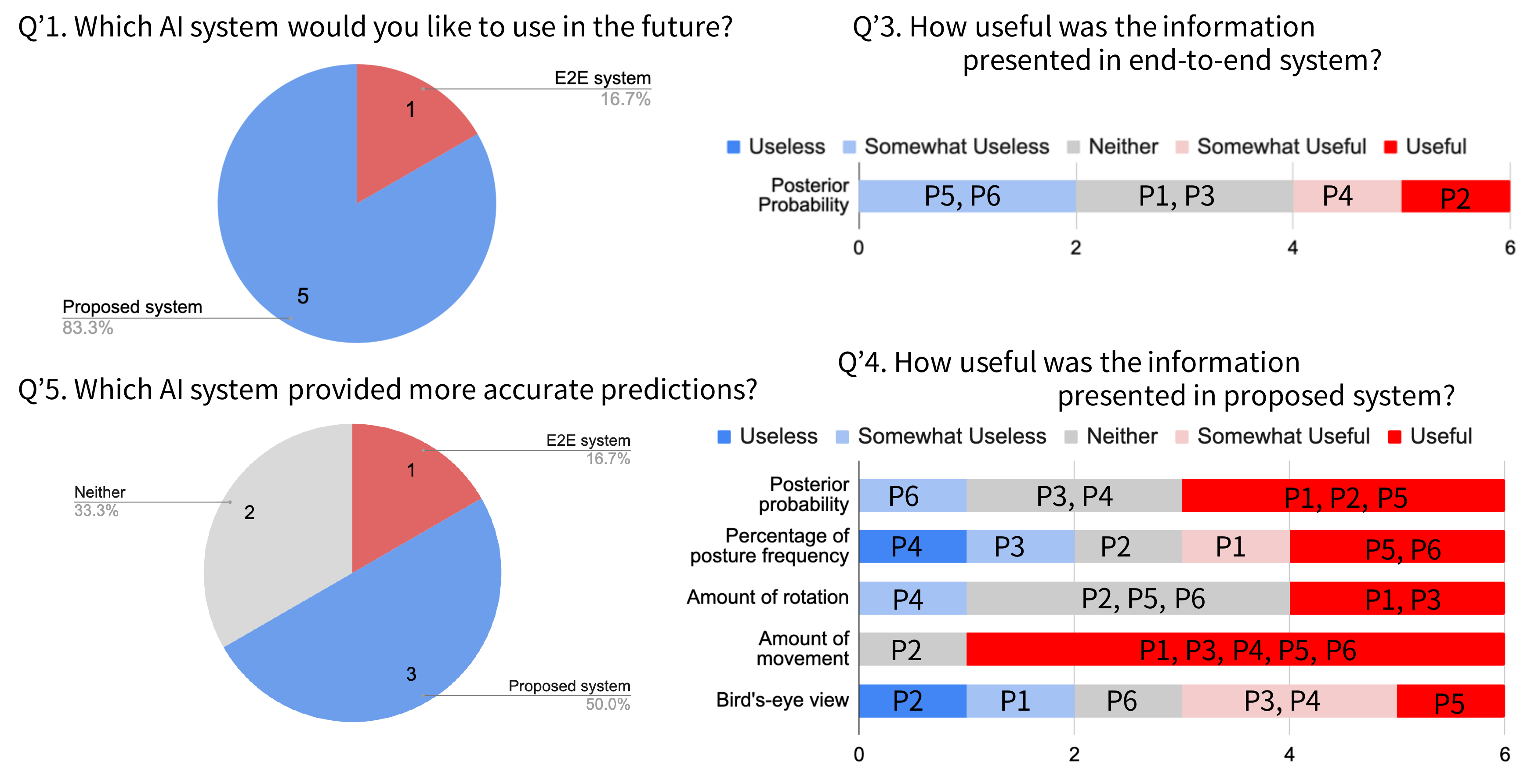}
    \caption{Responses to select questions in post-experiment survey.}
    \label{fig:results}
\end{figure*}

% 事後アンケートの結果
The results of the responses to the post-experimental survey are shown in Fig~\ref{fig:results}.
Five out of six participants selected the proposed system 
as the system they would like to use in the future~(Fig.~\ref{fig:results}, Q'1).
They found that the internal state of the system 
was helpful in identifying calving signs.
P1 said \textit{``It is easier to understand when the amount of movement, which is usually judged by the senses, is visualized in a graph.''}
P2 said \textit{``It's helpful to see detailed information about the cattle.''}
P3 said \textit{``The presented graphs were helpful in determining whether a behavior was a calving sign.''}
, and P5 said \textit{``Because the data is presented in detail, I think less experienced farmers can make decisions with more certainty.''}

In addition, 
some of the comments fit the hypothesis 
that the user interface of the end-to-end system,
which is a \textit{black-box},
is insufficient for decision-making.
P1 stated, 
\textit{``The [end-to-end system's] predicted probability alone does not tell us what we should do. I don't know what to do with it.''}.
In contrast,
P4, who was the only one to choose the end-to-end system in this question, 
said, \textit{``I did not find the indicators in the proposed system to be judgmental because there were many situations where the information presented differed from my perceptions.''}
P4 was a farmer who was asked 
to assist in the interviews to collect domain knowledge about calving signs,
and he stated that the information presented was less accurate than he expected. 
One shortcoming is that a user evaluation 
should have been conducted in the second phase of the framework design.
Because we did not receive sufficient feedback from experts in this phase,
the extracted features were not always sufficiently useful indicators.
In addition, 
it was suggested that actively presenting the internal state of the system, including the errors, may cause noise in judgment and cause participants to distrust the system.

Next, we discuss which information on the user interface was useful~(Fig.~\ref{fig:results}, Q'3, Q'4).
We found that the trend in the probability of calving signs varied between participants.
Only P4 responded that the probability of the end-to-end system was more useful, 
while the other participants found the proposed system to be more useful or about the same.
We now turn our attention to the results 
of the user interface of each statistic on the posture, rotation, and movement
of the proposed system~(Fig.~\ref{fig:results}, Q'4).
Two participants from the agricultural college (P5, P6) 
responded that the graph of the posture frequency was useful,
while the remaining four farmers responded that it was generally not useful.
This may be because the farmers could view the raw data 
to obtain more accurate information on posture frequency, 
so there was less of a need to refer to the information presented by the system.
On the other hand, 
participants from an agricultural college 
with an academic background 
found it useful to be able to quantitatively visualize the posture frequency.

A higher percentage of participants found the graph of the amount of movement to be more useful than the posture frequency graph because the meta-information, which cannot be captured from the raw data, is useful for judgment.
In fact, many of the participants said that the comparison with normal conditions was helpful during their responses.
P1 said
\textit{``Before, I judged the amount of rotation and movement intuitively, so it's easier to see when graphed.''}
and P6 said \textit{``The graph [of the amount of movement] makes it easier to notice the degree of change compared with the normal conditions.''}
The rotation graph tended to be less useful than the movement graph because the participants paid attention to the intensity of the cattle's movement as one of the criteria for judging whether it was a calving sign, and the amount of rotation was not directly related to their judgment. 
Thus, it is important to present features in a way that users can easily understand, and a more abstract expression such as ``intensity of movement'' may be more suitable.

% The participants indicated that 
% the predictive probability of the proposed system was somewhat more accurate.
In their responses to the last question (Q'7),
participants pointed out 
issues related to the accuracy of the presented information 
and that the user interface made the video smaller 
when information is presented. 
For the former issue, 
P4 said \textit{``I felt that the graph was not directly linked to the calving signs.''} 
and P6 said \textit{``I was worried when the AI identified an indeterminable posture in many cases. I would prefer to judge it myself in such situations.''}
For the latter issue, 
P5 stated, 
\textit{``Readability. I don't want the video to be too small.''}

% P6 said \textit{"I was worried about when the AI identified "Can't tell" posture in many cases. I thought it would be nice to be able to judge in such situations."}

%%------------------------------
%%------------------------------

\section{Discussion}
\label{sec:discussion}
% XAIのユーザ実験に関する一般化した議論
\textit{Black-box} AI systems are considered inadequate
for supporting human decision-making.
One farmer pointed out that the end-to-end system 
lacked instructions on what they should do 
after seeing the value for the probability of calving signs.
Several farmers pointed out 
the advantages of our proposed system, 
stating that the detailed information 
would help novice farmers make correct decisions confidently 
and the reason visualization was helpful. 
In particular, statistical information 
including meta-information that cannot be obtained 
from the raw data tended to be particularly useful. 
In contrast, 
we found that the reliability of the system is adversely affected 
when the presented information is incorrect.
This has been reported in several studies 
as algorithm aversion~\cite{Berkeley2014, Mary2002}, 
which is a phenomenon in which users stop trusting algorithms
after seeing its mistakes.
To present more information to users, 
the proposed framework needs to take into account 
the accuracy of the information.
In addition, 
we should have conducted a user evaluation 
in the second phase of the framework design
because the extracted features 
may not have been sufficiently useful indicators for users.
In the participants' feedback,
they noted that the user interface became more complicated
as more information was presented.
When considering development for actual use, 
it is necessary to consider elderly users 
as well as smartphones 
which may make the screen more difficult to view. 
A more sophisticated user interface design is needed, 
such as the separation of the summary and detailed analysis screens.
We anticipate that our evaluation of 
an XAI framework through user studies 
will contribute to the integration of XAI in various industrial domains.

\section{Conclusion}
\label{sec:conclusion}
% 結論
In this study, 
we proposed a framework for a video surveillance system 
that incorporates experts' decision-making processes
into the architecture such that 
the reasons for the prediction results can be interpreted.
We evaluated the calving sign detection systems 
based on our proposed framework 
through a user study with people involved in livestock farming (N=6).
The proposed framework was compared with an end-to-end system, 
and five out of six participants selected the proposed framework 
as the system they would like to use in the future. 
In addition, 
the proposed framework is used to present the internal state of the system, 
which can be used to help users make decisions 
and identify system errors. 
However, 
we found that presenting an inaccurate internal state of the system 
could interfere with the user's judgment 
and cause them to distrust the system. 
In future work, we intend to study 
the accuracy of the information presented to users.

\section*{acknowledgment}
We would like to thank the farmers who participated in the experiment 
and Kagoshima Prefecture Agricultural College.
We also thank Kagoshima Brain Center for fruitful discussions.
% We also thank Farmers Support Co., Ltd. 
% for setting up and assisting in the user experiments, 
% and Kagoshima Brain Center for fruitful discussions.

\bibliographystyle{IEEEtrans}
\bibliography{citation.bib}
\end{document}